\documentclass{article}

\usepackage{graphicx}    
\usepackage{amsmath}     
\usepackage{amsfonts}    
\usepackage{amssymb}     
\usepackage{hyperref}    
\usepackage{booktabs} 
\usepackage{natbib}
\usepackage{url,hyperref,lineno,microtype,subcaption}
\usepackage[onehalfspacing]{setspace}

\newcommand{\hl}[1]{#1}
\newcommand{\hlc}[2]{#1}

\begin{document}
\onecolumn

\title{Designing Interactions with Shared AVs in Complex Urban Mobility Scenarios} 



\author{
    Marius Hoggenmueller \\
    Design Lab, Sydney School of Architecture, Design and Planning \\
    \textit{The University of Sydney} \\
    \texttt{marius.hoggenmueller@sydney.edu.au}
    \and
    Martin Tomitsch \\
    Design Lab, Sydney School of Architecture, Design and Planning \\
    \textit{The University of Sydney} \\
    \texttt{martin.tomitsch@sydney.edu.au}
    \and
    Stewart Worrall\\
    Australian Centre for Field Robotics\\
    \textit{The University of Sydney}\\
    \texttt{stewart.worrall@sydney.edu.au}
}

\maketitle

\begin{abstract}
In this paper, we report on the design and evaluation of an external human-machine interface (eHMI) for a real autonomous vehicle (AV), developed to operate as a shared transport pod in a pedestrianised urban space. We present insights about our human-centred design process, which included testing initial concepts through a tangible toolkit and evaluating 360-degree recordings of a staged pick-up scenario in virtual reality. Our results indicate that in complex mobility scenarios, participants filter for critical eHMI messages; further, we found that implicit cues (i.e. pick-up manoeuvre and proximity to the rider) influence participants' experience and trust, while at the same time more explicit interaction modes are desired. This highlights the importance of considering interactions with shared AVs as a service more holistically, in order to develop knowledge about AV-pedestrian interactions in complex mobility scenarios that complements more targeted eHMI evaluations.

\tiny

\end{abstract}

\section{Introduction}

Fully autonomous vehicles (AVs) have the potential to not only mitigate accidents caused by human errors, but also fundamentally transform the way people commute in cities~\citep{Kellett2019}. Recent endeavours from government institutions and industry indicate a trend towards shared autonomous vehicles (SAVs) as a likely future mobility scenario, rather than people owning their personal vehicles~\citep{NARAYANAN2020255, Iclodean2020}. The promise of this approach is that the deployment of SAV services can have a positive impact on the quality of urban life, with less land being devoted to parking and less congestion. Models predict that the required fleet of SAVs to move the same number of people can be met with 70\% of the current taxi fleet for New York City and that the demand is equivalent to 30\% of the number of today's personal vehicles for Singapore~\citep{Pavone2015}.

The ubiquitous roll-out of AVs and SAVs is closely linked to overcoming technological challenges, such as sensing~\citep{Ilas2013}, in particular during poor lighting conditions~\citep{YONEDA2019253}, and optimising routing algorithms~\citep{LEVIN2017373}. At the same time, considering the human factors, including those affecting people outside the vehicle, has gained attention from industry and academia~\citep{Mora2020}. For example, there is an increasing body of work investigating the use of external human-machine interfaces~(eHMIs) to overcome the challenge of how AVs can communicate their internal state to nearby pedestrians. Examples range from projections on the street~\citep{Nguyen2019} to using light strips attached to the vehicle~\citep{DeyChi2020, Eisma2020}. A recent literature review by~\citet{Dey2020taming} found that the majority of concepts only focuses on communicating information related to the vehicle's yielding intent (i.e. whether it is safe for other road users to cross in front of a vehicle); further, concepts which have been evaluated through empirical studies mainly cover simplistic traffic scenarios, for example, one person crossing a roadway in front of an AV~\citep{Colley2020chi}. This indicates that there remain several unresolved questions when it comes to designing interactions between SAVs and pedestrians that are not addressed by previous eHMI concepts and empirical studies. Many open questions remain, such as whether an eHMI is able to successfully encode information that is broadcast to the general public (e.g. a vehicle's intention and awareness) while at the same time showing information relevant to a particular rider (e.g. to identify which SAV is theirs). Further, with the roll-out of SAVs as a last-mile transport mode between larger hubs, such as train stations, and the passengers' final destination~\citep{Yap2016}, it is likely that those vehicles will operate in pedestrianised areas rather than on dedicated roads. A government report published by one of Australia's transport authorities noted that research on pedestrian safety in shared spaces is widely underrepresented~\citep{NSWReport}, which echoes the systematic review by~\citet{Dey2020taming}, finding that eHMI studies mainly focus on intersections and crossings.

In this paper, we report on findings from a research project that involved designing a low-resolution lighting-based eHMI for a shared passenger transport pod. Following a toolkit-supported human-centred design process, we developed an eHMI to display the vehicle's status, intent, and awareness, as well as to enable users to identify their vehicle. To evaluate the eHMI, we devised a ride-sharing scenario with multiple vehicles commuting in a shared urban environment where pedestrians, cyclists, and maintenance vehicles share the same road. The scenario was captured with a 360-degree video camera and represented to participants (N=14) in a virtual reality (VR) environment. Through this study setup and feedback collected from participants via semi-structured interviews, we investigated the efficacy of eHMI communication in complex urban mobility scenarios. We specifically focused on three aims: The use of eHMIs to convey multiple messages simultaneously, participants' perception of multiple AVs and their eHMIs, and AV-pedestrian interactions for SAVs in a shared space.

The paper contributes to the field of automotive user interfaces broadly and to AV-pedestrian interaction specifically in two ways. First, it offers insights about the role of implicit (e.g. vehicle behaviour) and explicit (e.g. eHMI) cues and how people perceive those cues in different scenarios (e.g. crossing versus pick-up). Second, it provides an account of human-centred methods and their value for designing AV-pedestrian interaction in complex scenarios.

\section{Related Work}

\subsection{AV-Pedestrian Interfaces}
In recent years, researchers have stressed that autonomous vehicles require additional means to communicate to other road users~\citep{Mahadevan2018, Rasouli2020}. Due to the absence of a human driver, interpersonal communication (e.g. eye contact or gestures) and the manual use of signalling devices (e.g. indicators, horn) are not longer available. However, researchers stressed that such communication cues are important, in particular in dense urban areas, where vehicles share spaces with vulnerable road users (e.g. pedestrians)~\citep{hollaenderChi2021} and right-of-way negotiation is necessary. As a consequence of addressing this issue, there exists now a growing body of work on external human-machine interfaces (eHMIs)~\citep{Dey2020taming}. Concepts range from projection-based eHMIs~\citep{Nguyen2019} to such attached to the vehicle itself, for example light band eHMIs~\citep{DeyChi2020}. In right-of-way negotiations~\citep{Clercq2019}, most of the eHMI concepts incorporate the vehicle's yielding intent~\citep{DeyChi2020}. While there has been research suggesting that pedestrians mainly inform their crossing decision based on implicit cues, such as motion~\citep{Moore2019, Dey2017, Risto2017}, other empirical studies have shown that status+intent eHMIs can significantly reduce the risk of collisions with AVs~\citep{Faas2021} and increase pedestrians' subjective feeling of safety~\citep{Hollaender2019}. Other research on eHMIs has studied interface placement on the vehicle~\citep{Eisma2020}, communication modalities (e.g. light band eHMIs for abstract representations~\citep{DeyChi2020}, or higher resolution displays for text and symbols~\citep{Hollaender2019, Chang2017}), as well as message perspective~\citep{Eisma2021}. Furthermore, researchers began to investigate external communication concepts beyond crossing scenarios: for example, \citet{Colley2020a} investigated the specific situation in which automated delivery trucks would block parts of the road and sidewalks and designed and evaluated a visualisation concept that guides pedestrians to safely walk past the truck. 
Others conceptualised autonomous vehicles as public displays that can do more than display information related to the vehicle's operational task and pedestrian safety, such as showing navigation cues and advertisements~\citep{Colley2017, Asha2020, Colley2018perdis}. However, despite the plethora of eHMI concepts, systematic reviews~\citep{Dey2020taming, Colley2020b} have emphasised that a majority of design concepts are limited to one specific traffic situation, mostly uncontrolled zebra crossings, and only few empirical evaluations take into account urban contexts beyond the road, such as shared spaces~\citep{Yang2021}.

\subsection{Shared Autonomous Vehicles}
The global rise of ride-sharing services (e.g. Uber) and the expected uptake of SAVs has led to growing interest from the human-computer interaction (HCI) community~\citep{Eden2017}. Researchers began to systematically study aspects that influence passenger's experience and trust towards those services, including trip planning~\citep{Svangren2018}, and how to design for in-vehicle experiences~\citep{Khamissi2019, Braun2018}, for example, informing passengers about their current trip~\citep{Flohr2020} or communicating the vehicle's driving decisions~\citep{Sandhaus2018}. Researchers have also identified potential security concerns of sharing AVs with others~\citep{Schuss2021} and explored the needs of specific user groups, such as the elderly~\citep{Gluck2020} or children~\citep{Kim2019}, with the aim to design for more inclusive in-vehicle experiences. 

On the other hand, passenger's experience with SAV services in situations outside the vehicle (e.g. while waiting for an approaching vehicle) has received little attention so far. To the best of our knowledge, only \citet{Florentine2016} and \citet{Verma2019a} developed design concepts for eHMIs on SAVs, but those only focused on displaying intent, did not specifically address a passenger-pedestrian perspective, and were evaluated in crossing situations only. \citet{Owensby2018} developed a framework for designing interactions between pedestrians and autonomous vehicles in more complex scenarios. They used a ride-sharing scenario as a foundation for developing and validating the framework. Building on the work from \citet{Robertson2009} on designing situations, the first proposed step is to break down the scenario into different stages (used synonymously for situations that unfold in an AV-pedestrian interaction scenario). Those stages are then mapped onto three high-level dimensions addressed for each specific situation: how information is being presented, the interactions between user and system, and the user needs being addressed. While the framework is a good starting point (and indeed provided us with the conceptual foundation for our own design process), it has not previously been applied or validated in a larger study. Using the framework as a foundation, in this paper, we designed a comprehensive and consistent set of eHMI visualisations for a shared AV and evaluated those in a contextualised study setup (i.e. an immersive VR environment~\citep{Flohr2020}).

\section{Design Process}
In this section, we report on the iterative process of designing the eHMI for an autonomous transport pod \hlc{as part of an interdisciplinary research project. The project team involved robotic engineers (referred to as ``engineering team'' in this section), interaction designers (referred to as ``design team'') and urban planners. During the 8 months design process (i.e. from initial discussions up to the completion of the VR prototype), we had regular internal planning meetings approximately once every two weeks. In the meetings, the larger team provided feedback to the design team on the eHMI light pattern iterations and planned further research activities, such as the design exploration sessions with external experts. The urban planners provided targeted advice on the chosen urban context and scenario.}{more details on the overall design process (R1 / R3)} Below we describe the a) chosen urban context, scenario, and unfolding situations that the eHMI was designed for, b) the hardware setup, c) the design of the eHMI concept, which was informed by toolkit-supported collaborative design exploration sessions with external experts, and d) the VR prototype, which was used to evaluate the scenario and eHMI with potential users. 

\subsection{Urban Context, Scenario and Situations}

As the study used an existing, fully functional AV, we selected an urban context that suited the operational specifications of the vehicle. The AV was developed as a pod rather than a full-scale car, allowing it to operate in shared spaces. The engineering team had been granted permission to operate the AV on our university's campus, which resembles a shared space, as our campus avenues are frequented by pedestrians, cyclists, and authorised vehicles (e.g. for delivery or maintenance). Thus, we situated our AV-pedestrian interaction scenario on one of our university's main avenues with no road markings and a consistent amount of pedestrian traffic. As a specific scenario, we chose a passenger pick-up scenario given the likely role that SAVs will play in future mobility implementations~\citep{Schuss2021}. SAVs have further been implemented on less traveled routes, such as University campuses, already~\citep{Iclodean2020}. Choosing this scenario also allowed our study participants to draw on their previous experience with ride-sharing services, such as Uber. 

The scenario further allowed us to map out and design how the eHMI would support AV-pedestrian interaction for a number of specific situations. In other words, we broke down the complex urban scenario of interacting with multiple ride-sharing vehicles in a shared space into a set of situations. Specifically, we identified four situations, using the framework by \citet{Owensby2018}, which outlines interactions in an autonomous ride-sharing scenario. The situations involved (1) an SAV driving along the shared avenue, (2) the SAV pulling over to pick up a rider, (3) the rider boarding the SAV, and (4) a pedestrian crossing in front of an SAV (in order to illustrate that the vehicle is aware of surrounding people). \hlc{The chosen situations required us to address the four user requirements previously identified by \citet{Owensby2018} for autonomous ride-sharing scenarios, namely (1) being able to identify the vehicle, (2) knowing the current status of the vehicle, (3) knowing the vehicle's intent, and (4) that the vehicle is aware of the user~(the rider and surrounding pedestrians).}{User requirements which guided the eHMI design process (R2)}

\begin{figure}[!t]
\centering
\includegraphics[width=0.5\linewidth]{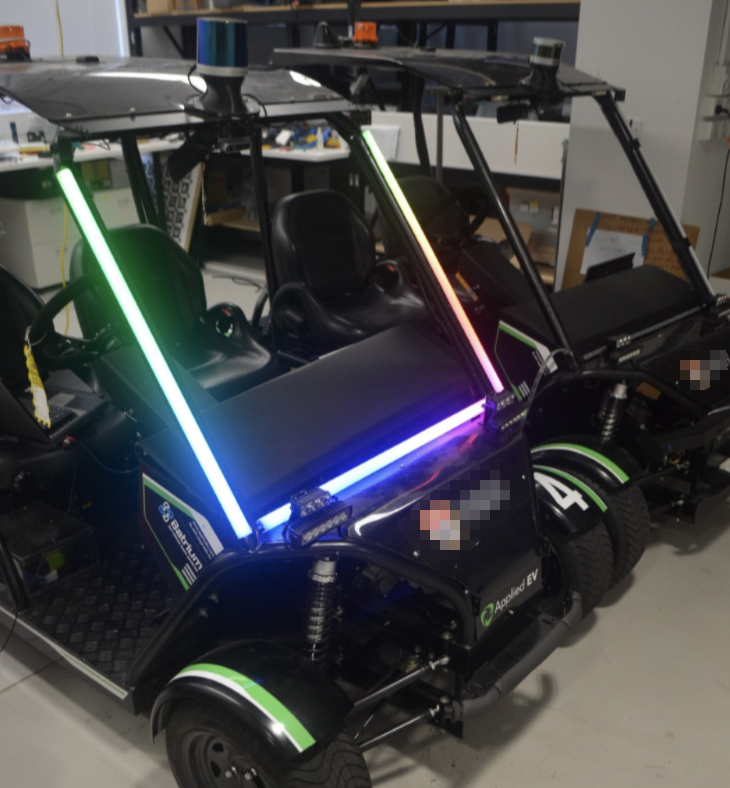}
\caption{Passenger transport pod with "U"-shaped low-res lighting display.}
\label{fig:vehicle}
\end{figure}

\subsection{Passenger Transport Pod and eHMI Hardware} 

We designed the eHMI visualisations for a fully functional AV passenger transport pod, which was also used later on for the recording of the immersive 360-degree prototype. The AV hardware was designed by AEV Robotics\footnote{\url{https://www.instagram.com/aevrobotics/}, last accessed: January 2022} and was further customised by our engineering team. The platforms have the sensing and computation capacity to eventually operate at SAE level 5\footnote{The automation levels are defined by the Society for Automotive Engineers (SAE) for autonomous driving. Level 5 refers to full automation: \url{https://www.sae.org/news/2019/01/sae-updates-j3016-automated-driving-graphic}, last accessed: January 2022} and are based on the robot operating system (ROS). The vehicles -- being small, efficient and electrically powered -- were designed for the purpose to operate safely in low speed road environments (under 40 kph). This makes them suitable to operate in close proximity to pedestrians~\citep{Pavone2015}. One single vehicle is intended to carry up to two passengers.

The engineering team decided early on to use an LED-based low-resolution (low-res) lighting display to implement the final eHMI. This decision was made due to the relatively low power consumption of LEDs, thus being able to power the eHMI with the vehicle's on-board battery. Furthermore, LED light strips are a widely available technology which makes it easy to apply this eHMI solution to similar AV platforms~\citep{Dey2020taming}. Low-res lighting displays have been previously studied in pervasive display research as they allow to communicate information at the periphery of attention~\citep{Offenhuber2014} and can be perceived from a distance in outdoor environments~\citep{Wiethoff2017}. For this reasons, low-res lighting displays have been also widely used for the implementation of eHMIs in crossing scenarios (e.g.~\citet{Verma2019, DeyChi2020}), and previous research has indicated that simple visual cues are easy to understand also in particular for child pedestrians~\citep{Charisi2017}. 

The engineering team installed off-the-shelf LED strips\footnote{\url{https://www.pololu.com/category/180/sk6812-ws2812b-based-led-strips}, last accessed: January 2022} around the front window of the vehicle in a "U"-shape (see Figure~\ref{fig:vehicle}).
The LED strips featured a pitch of 60 pixels per meter, resulting in a total of 145 LEDs. The LEDs were controlled via an Arduino board, which was connected to the system of the vehicle. A python ROS node read the information from the vehicle state by subscribing to the relevant information. Light patterns were triggered in real-time based on the sensed information~(awareness) and the state of the AV platform~(intent). \hlc{After conducting several tests in the real-world and under different lighting conditions, the designers advised the engineering team to install a diffuser tube of opal white acrylic wrapped around the LEDs. Following design recommendations for low-res lighting displays~\citep{Hoggenmueller2018}, this decision was made to improve the viewing angle and to create the illusion of a light bar (rather than a distinct set of point light sources). At this stage of the design process, we also took into account the subsequent production of the virtual reality prototype using a 360-degree camera (see section 3.4). For this particular purpose, adding the diffuser tubes significantly improved the visibility of the eHMI and eliminated the glaring effect in the recordings that we observed when capturing the LEDs without the diffuser tubes.}{additional information on the design process and collaboration between design \& engineering (R1)}

\subsection{Designing eHMI Light Patterns}
Designing the eHMI light patterns for the low-res lighting display, the design team followed an iterative design process, which involved the use of a tangible toolkit for prototyping AV-pedestrian interactions~\citep{hoggenmueller2020ozchi}. The toolkit was used to (a) quickly prototype different visualisation concepts, (b) present concepts during \hl{internal} team meetings in a more tangible manner, and (c) to facilitate collaborative design exploration sessions \hlc{with recruited expert participants to further inform the design of the eHMI light patterns}{clarifying that the workshop participants were recruited external experts (R1)}. 
Below we describe the key features of the prototyping toolkit, the results from seven expert workshops and the final set of eHMI light patterns.

\subsubsection{Tangible Multi-Display Toolkit}
\begin{figure}
\centering
  \includegraphics[width=0.6\linewidth]{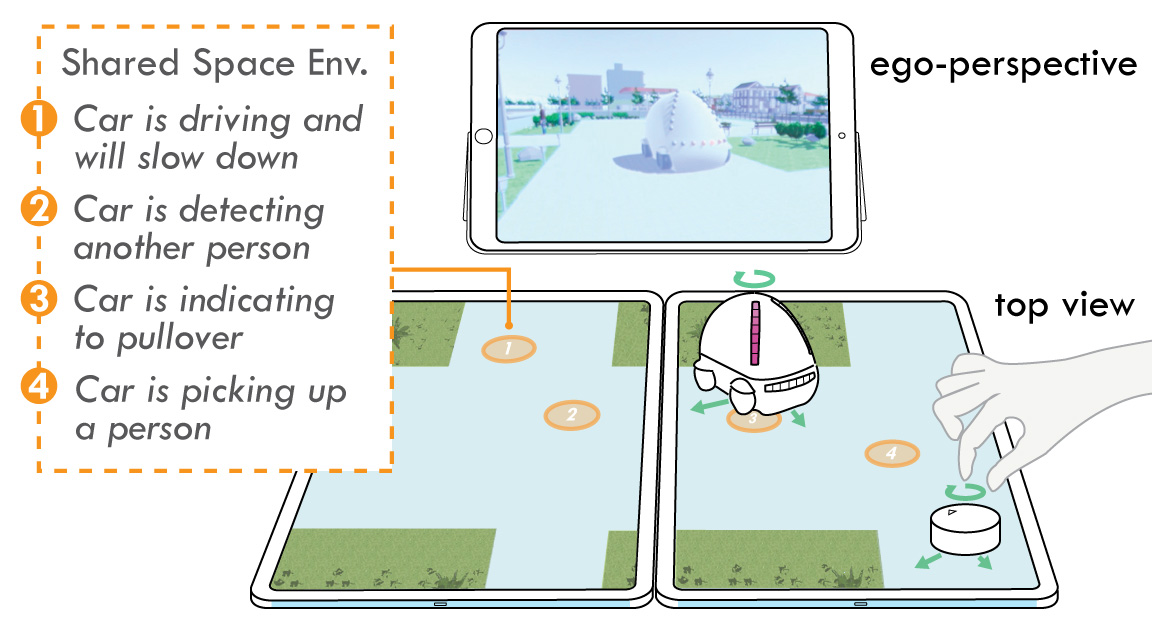}
  \caption{The tangible multi-display toolkit used to inform the eHMI visualisation design via computer simulations across multiple displays to capture different viewing angles, tangible objects to interact with the simulated environment and to depict the eHMI’s behaviour through an integrated miniature LED display.}
  \label{fig:toolkit}
\end{figure}

Building on small-scale scenario prototyping techniques~\citep{Pettersson2017} tailored to the context of AV-pedestrian interfaces, a toolkit approach was used to inform the eHMI visualisation design (Figure~\ref{fig:toolkit}). The toolkit enables multiple viewing angles and perspectives to be captured simultaneously (e.g. top-view, first-person pedestrian view) through computer-generated simulations orchestrated across multiple displays. Users are able to directly interact with the simulated environment through tangibles, which physically simulate the interface’s behaviour (in our case through an integrated LED display). Furthermore, a configuration app running on a separate tablet allows to control and adjust the design options in real-time. For the purpose of our project, this allowed users to change between various light patterns and adjust colour schemes and animation speed.

\subsubsection{Expert Workshops}
We conducted seven workshop sessions in total, with each session involving a pair of \hl{external }expert participants. The aim of the workshops was to receive feedback on light pattern candidates and identify a final set for further implementation on the AV. We recruited 14 participants (seven male, seven female) of various academic and professional backgrounds, covering a range of expertise considered relevant for the design of urban technologies~\citep{Tomitsch2020, Malizia}. Their areas of expertise included architecture and urban planning (n=5), human-computer interaction (n=5), psychology (n=2), software engineering (n=1) and civil engineering (n=1). Each workshop session lasted 90 minutes in total and was video-recorded for later analysis. \hlc{Having participant pairs allowed the experts to have more natural conversations with each other~\citep{Nielsen1993}. This co-participation setup has further been found to be preferred by participants and to detect a higher number of usability issues when evaluating design proposals~\citep{Mayhew2018}. Participants were randomly paired.}{Justify study setup (R1)}

In preparation of the workshop we implemented 12 different light patterns for our four AV-pedestrian situations (i.e. three pattern candidates per situation) for a ride-sharing scenario with the toolkit. The design of the light patterns was informed by previous \hlc{eHMI research~\citep{Nguyen2019, Mahadevan2018, Florentine2016,Dey2018,boeckle2017}}{Adding missing refs (R1)} and went through several iterations based on internal discussions within the project team: \hlc{for example, at the beginning of the design process, we considered re-purposing the SAV's existing front lights to indicate the intent to pull over. However, we rejected this idea later and opted for an eHMI solution that would integrate all messages in the same display space. This decision was made for aesthetic purposes but also in regards to the emerging research question whether a single low-res display would be capable to successfully communicate multiple eHMI messages~\citep{Dey2020taming}. Considering related literature on ambient light systems~\citep{Matviienko2015}, we applied different information encoding parameters (e.g. colour, brightness, LED position, or combinations thereof) for the different light pattern candidates.
For example, for the situation of the vehicle slowly moving in autonomous mode, we designed a purely colour-based pattern to indicate low speed, a pattern encoding slow speed through the size of the light bar (i.e. the numbers of adjacent LEDs lighting up), and a pulsing pattern changing the brightness at a low frequency.}{extending on design iterations (R1)}
Participants were presented with each of the 12 light patterns and asked to interpret their meaning and  
to provide feedback on the eHMI visualisation design. Participants were encouraged to make changes to the colour schemes via the configuration app as part of their design exploration. At the end of each workshop, we asked participants to select their preferred set of light patterns across all four situations. 

\subsubsection{Final Set of Light Patterns}

Based on the analysis of the participant input collected during the workshops, we derived several insights that guided our subsequent design decisions. These included: avoiding the use of red and green colours, using subtle light patterns by default (in regards to the shared space context in which pedestrians have right of way), using strong signals only when the car is going to do something unexpected or in high-risk situations, using a light pattern that is distinct from a turn signal when pulling over to pick-up a rider, and using a subtle animation for indicating the rider to get on the car (to avoid that the rider feels rushed or distracted during the boarding process). \hlc{In particular, the use of red and green colours to indicate the vehicle's speed caused confusion or different opinions among our workshop participants. While the majority of participants could establish a connection to the vehicle's speed, some participants interpreted the colours the opposite of our intention to encode low speeds through green colours, reversibly using red at high speeds. Even those participants who interpreted the colours correctly expressed their concerns about the potential ambiguity of this approach.}{more details on the findings from the expert workshops (R1)} Furthermore, encoding the vehicle's current speed and speed intent (acceleration/deceleration) was deemed as rather not relevant in a consistent low-speed environment. Instead, participants preferred the AV to signal that it is operating in a low-speed autonomous driving mode to express that the vehicle is always aware of its surroundings. 

\begin{figure*}
  \includegraphics[width=\textwidth]{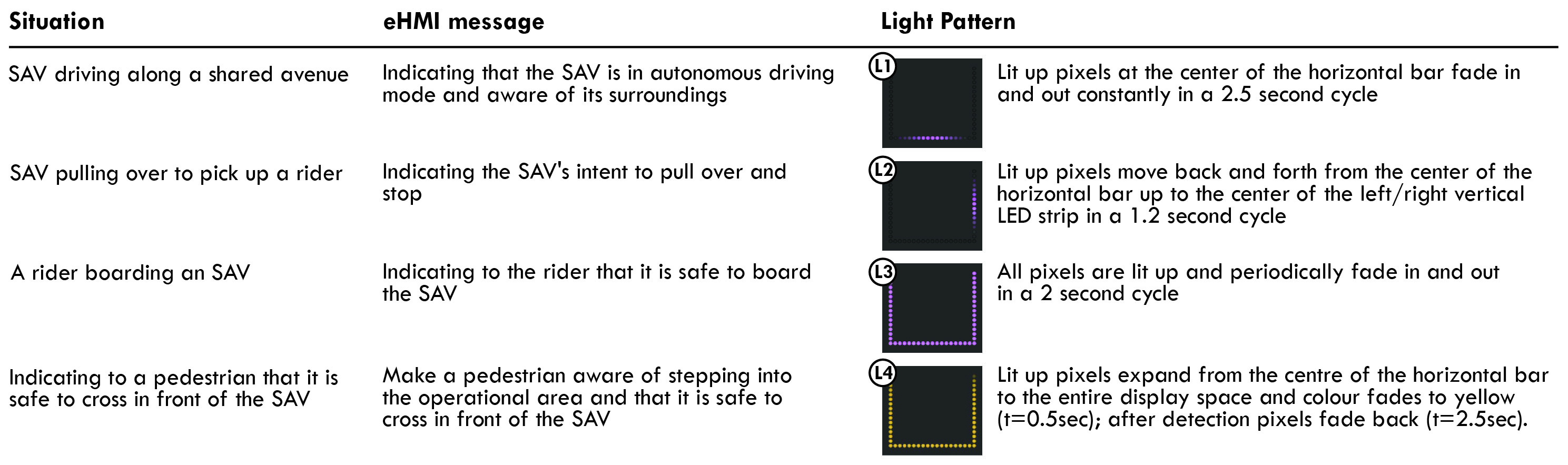}
  \caption{Overview of the encoded eHMI messages and light patterns linked to the previously identified situations. Light pattern L1-L3 are in purple colour which we used in the VR study for participants to indicate their vehicle.}
  \label{fig:visualisations}
\end{figure*}

Following the workshops, the design team revised the light patterns and implemented them as a simulation in Adobe After Effects. These simulations were then passed onto the engineering team along with a specification document for each pattern. The engineering team implemented the patterns as eHMI visualisations for the AV, to allow for further testing in a real-world context. The final eHMI visualisations and light patterns are depicted in Figure \ref{fig:visualisations}. As colour was deemed the most intuitive way for a low-res lighting display to represent the identification of the vehicle, we decided to represent the remaining messages (status, intent) through animation patterns only. \hl{This design decision was also confirmed through the feedback from workshop participants who mostly considered LED position and animations sufficient to encode those messages and suggested to avoid the use of red and green colours related to the vehicle's status.} Thus, the identification of the vehicle through colour is laid on top of the other cues. If the vehicle is not intending to pick up a rider, the animation pattern is displayed in a more neutral white. Only for the awareness cue, we decided to use a yellow colour in order to add further emphasis on the potential safety hazard through an additional change in colour. We decided for yellow as a more neutral colour compared to red (i.e. as previously suggested by ~\citep{DeyChi2020} for eHMIs), and this was also confirmed by some of our \hl{workshop participants}, who associated red with a potential malfunctioning of the vehicle.

\subsection{Virtual Reality Prototype}

To safely test the eHMI visualisations in a real context, we opted for creating a 360-degree virtual reality (VR) prototype representation. This kind of prototype, also referred to as hyperreal prototype~\citep{Hoggenmueller2019hyper}, has been found to result in an increased sense of familiarity in participants~\citep{Gerber2019} compared to other representations, such as computer-generated VR prototypes. VR was chosen over a field study to reduce any potential risk for study participants and as it is a commonly used approach for evaluating AVs and their eHMIs~\citep{Shuchisnigdha2017}. Using a pre-recorded video prototype further enabled us to test the situations under the exact same conditions across participants, thus balancing ecological validity and reproducibility of the study findings. 

\begin{figure*}
  \includegraphics[width=\textwidth]{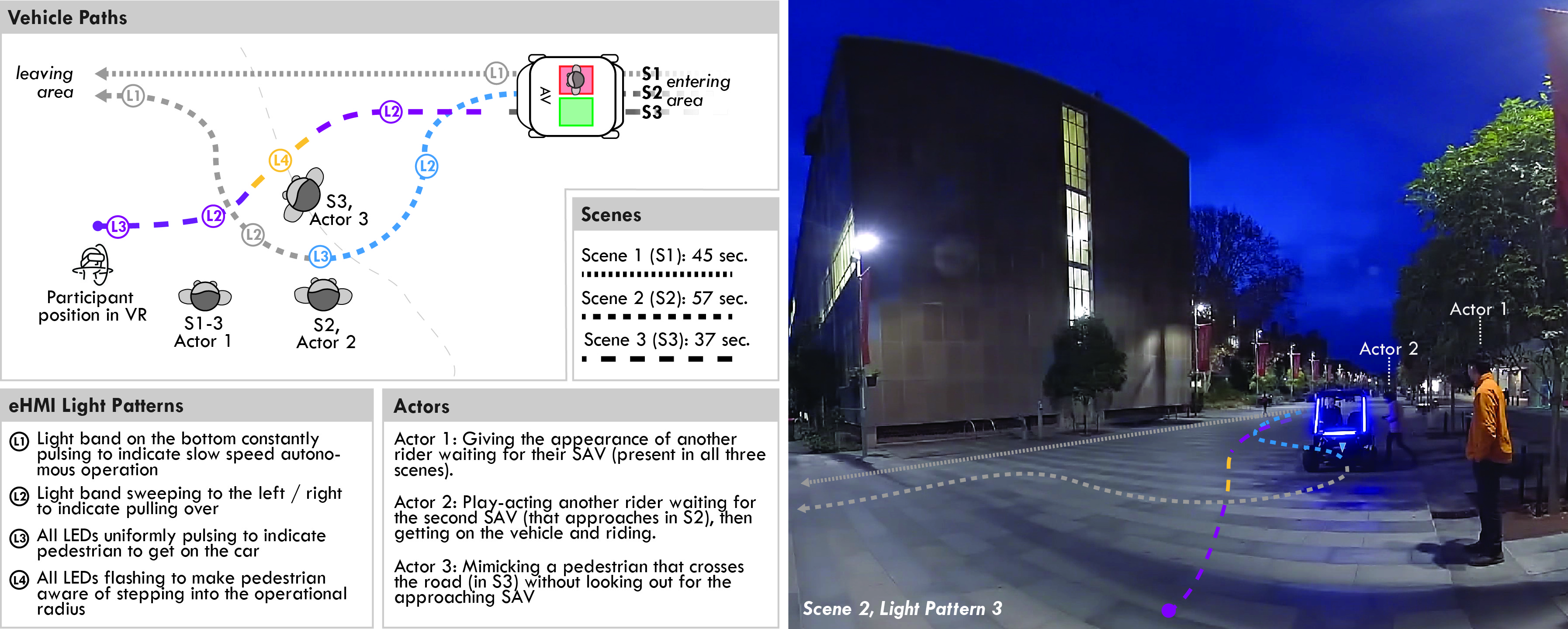}
  \caption{Recording plan of the three scenes, vehicle trajectories and eHMI light patterns (left); screenshot taken from the 360-degree video prototype representing the second scene with an actor entering the SAV (right).}
  \label{fig:recording}
\end{figure*}

We started with creating storyboards to capture the staged situations and interactions, which involved four actors to represent a pedestrian crossing in front of the SAV, a person boarding an SAV, a person waiting for their SAV to arrive, and a rider inside the SAV. 
We decided to spread our four situations
over three consecutive scenes (Figure~\ref{fig:recording}). This decision was made for two reasons: firstly, all staged AV-pedestrian interactions had to occur not too far away from the camera stand for later visibility in VR; secondly, we wanted to give participants the impression that multiple SAVs are commuting through the shared space rather than a single one, however we only had one eHMI-equippped AV available. The scenes (represented from the perspective of the study participant) included: (1) The SAV passing through the shared environment without any staged interactions with pedestrians, (2) the SAV pulling over and picking up another rider (Actor 2 in Figure~\ref{fig:recording}), and (3) the SAV indicating to pull over to the camera stand. In the third scene, a pedestrian (Actor 3 in Figure~\ref{fig:recording}) crosses in front of the SAV, forcing it to slow down and stop. 
An additional person was placed directly behind the camera in all three scenes (Actor 1 in Figure~\ref{fig:recording}), giving the appearance of another rider waiting for their SAV. This was to constrain participants' movement in the simulation, as 360-degree video does not allow for motion when imported into VR.

We did several tests of the SAV's behaviour within the real urban context and to prepare the AV for recording the scenes. At the time, the AV had been programmed to use a combination of algorithms and a cost map that kept the vehicle as close to the middle of the avenue as possible.
Upon testing the SAV's behaviour when approaching a rider, we found that the SAV would move in a straight line towards the waiting rider, which was in conflict with previous observations that AVs should mirror the behaviour of human drivers~\citep{Schneemann2016}. Through informal tests with members of the project team, we also found the direct approach to be perceived as threatening from the perspective of the waiting rider. 
Hence, we programmed the SAV to follow a pathway that was recorded based on a human driver pulling over to the side of the avenue following an S-curve trajectory. 
On top of the prerecorded trajectory, the vehicle was operating a `virtual bumper' which is a system that detects obstacles in (or adjacent to) the proposed vehicle trajectory and reduces the speed based on a time-to-collision calculation. Due to safety regulations, a licensed operator had to sit in the SAV -- in case of having to manually bring the SAV to a halt. However, for the purpose of the recordings, we were able to remove the steering wheel, thus conveying clearly to participants that the vehicle was operating autonomously with the operator playing the role of a rider. 
The eHMI light patterns were fully implemented and connected with the SAV's operating system and programmed to respond with the appropriate message for each of the staged situations. 

We recorded the scenes in the chosen urban context, a pedestrianised area on our university campus which leads to the university’s main buildings. We used an Insta360 Pro 2\footnote{\url{https://www.insta360.com/product/insta360-pro}, last accessed January 2022} camera (capable of recording 360-degree panorama videos in 8K 3D). For post-processing purposes we used Adobe Premiere and Adobe After Effects. As we recorded the scenes during dusk for better visibility of the low-res lighting display, we had to apply the Neat Video\footnote{\url{https://www.neatvideo.com/}, last accessed January 2022} filter to reduce image noise, while still preserving fine details, such as people's faces. We then combined the three scenes, added a short blend transition between them, and exported them into a single 3D over-under video file. To experience the stereoscopic 3D 360-degree video with a VR headset (HTC Vive), we imported the video file into Unity and applied it as a render texture on a skybox material. 

\section{Evaluation Study}

\subsection{Materials and Setup}

The study took place in our VR lab space (approx. six by six metres). We used an HTC Vive VR headset for the experiment. To convey the immersive audio recording of the scene soundscape and increase a sense of presence, we used stereo headphones. We further prepared a mock-up interface for a mobile SAV ride-sharing application which showed the following information: (a) a map of the location where the participants were supposed to wait for their vehicle, (b) the vehicle’s current position approximately two minutes away from the participant, (c) the colour which was assigned by the system for the participant to recognise their vehicle (in this case purple), and (d) a mock user profile of the other rider whom they would share the approaching SAV with. 

\subsection{Participants and Procedure}
The study involved 14 participants (seven male, seven female). None of those participants had been involved in the expert workshops. Ages of the participants ranged between 21 and 55 years (\textit{M=31.42}, \textit{SD=8.6}). Out of our participants, six were students and eight working professionals; three participants never experienced VR before, eight participants had less than five experiences in VR, and three participants more than five. We recruited participants from our university's mailing lists, flyers, and social networks. Taking part in the study was entirely voluntary and initial contact had to be made by the participants, following the study protocol approved by our university's human research ethics committee.

After arriving in our lab, we first gave a short introduction to each participant about our research and informed them about the study purpose of evaluating interactions between SAVs and surrounding pedestrians (including waiting passengers). Each participant filled out the study consent form and a short questionnaire to collect demographic data. We then quickly briefed participants about the designed scenario of waiting for a requested SAV service. Before commencing with the VR experience, we presented them with the mock-up interface of the SAV ride-sharing application. The duration of the scenario in VR was 2 minutes and 19 seconds. The duration was chosen based on previous tests with members of the wider project team, ensuring that the scenario was long enough for participants to be immersed in the scenario, but at the same time short enough to avoid study fatigue. After experiencing the scenario in VR, each participant partook in a post-scenario semi-structured interview.

\hlc{Out of the 14 participants, three reported that they had experienced VR more than five times before this study, eight reported that they had experienced VR at least once but less than five times, and three reported that they had no prior VR experience. Interestingly, while the majority of our participants had previous experience in VR, none of them had experienced 360-degree videos in VR before but only computer-generated content. One participant (P10), for example, stated: \textit{`Previously, what I was used to in VR was like a game, so it was not necessarily a realistic situation.'} Potentially as a consequence of this, participants commended the high visual realism of the VR experience.}{adding information on prior VR experience and perceived realism of VR experience (R3)}

\subsection{Data Collection and Analysis}

The post-scenario semi-structured interview included questions covering three broader areas: (a) participants' perception and understanding of the eHMI, (b) participants' trust towards the vehicle, and (c) their general experience of the SAV service, all based on the scenario which they experienced in VR. The interview took 8 minutes 39 seconds on average (SD=3 minutes 36 seconds). The interviews were audio-recorded for later analysis. Additionally, we also took notes about participants' behaviour when experiencing the VR prototype (e.g. if participants made comments or gesticulated during the experience).

The interviews were transcribed by a professional transcription service. Two researchers were involved in coding the transcribed interviews, following the thematic analysis approach~\citep{BraCla06}. The two coders started the coding process with a different set of interviews. Later on they used a collaborative online whiteboard to look for agreement and disagreement between their codes and to develop the final categories and overarching themes.

\subsection{Results}

The results are structured following the themes that we conceptualised through the thematic analysis of the interview data. Where relevant we augment the findings with observations recorded during the VR experience. 

\subsubsection{Interpretation of the eHMI}
In our study, participants experienced three scenes accommodating various traffic situations and eHMI messages. In the post-scenario interviews, when being asked about the light patterns, participants most frequently referred back to the eHMI \hl{light pattern L4} that would make other pedestrians aware of stepping into the operational radius of the vehicle (n=10), and the \hl{colour encoding (i.e. purple)} that would help participants to identify their approaching vehicle (n=10). Only one participant (P10) mentioned the eHMI \hl{light pattern} of pulsing white colours (L1) when the vehicle was just commuting through the shared space and signalling its autonomous operation mode. P10 stated: \textit{`To me it was clear that that wasn’t my car, so I sort of looked at it but I just ignored it.'} Also, participants often did not discern between the different sequential eHMI messages (i.e. pulling over, signalling to get on the car) when their vehicle was approaching. Five participants (P4, P7, P9, P12) stated explicitly that they only focused on the colour for identifying their SAV. For example, P7 said that \textit{`[she] was just thinking about matching'}, and \textit{`didn’t interpret the light patterns as any kind of indication of movement or intent'}. Similarly, P4 stated that \textit{`[she] was just looking at the colours, [...] and wasn’t expecting any other meaning from the display'}. In a similar vein, P9 stated that \textit{`[he] was just trying to concentrate on which was [his] vehicle, and [he] didn't look for any additional information'}. Participants who recalled the animation patterns in the interviews expressed mixed opinions. For example, P7 stated that the sweeping animation to indicate pulling over (L2) \textit{`is more intuitive [...] as it conveyed the directionality better than just the on and off [i.e. referring to a conventional blinker]}. On the other hand, P13 found the sweeping animation \textit{`way too abstract'}, similar to P10 referring to it as \textit{'fancy indicator'} and P8 who stated: \textit{`That’s a massive [...] change, if you’re now saying a car’s indicator is not an indicator anymore, whereas it’s been like that for a century'}. Here, a common concern was also that the pulling over animation was functionally and spatially overlaid with the light pattern that helped participants to identify their vehicle.
The majority of participants did not make similar comments about the \hl{animated light pattern} indicating participants to enter the car (L3) or the \hl{light pattern indicating alert to pedestrians when stepping into the AV's operational radius} (L4).

\subsubsection{Colour Differentiation and Multiple Vehicles}
In our prototype, we deliberately decided for two similar, yet distinguishable colour codes for identifying the vehicle, namely blue (hex colour code:~\#46CCFF) for Actor 2 and purple (\#876AE8) for the VR participant.
While many participants (n=10) recalled on the colour code to identify their vehicle, all but one participants also raised that they experienced difficulties in confidently identifying their vehicle based on the assigned colour. For example, P8 stated that he \textit{`couldn’t distinguish the major difference between those two colours'}. P7 highlighted the limitation of using colour to encode important information in terms of the difficulties this would create for colourblind people. Eight participants explicitly brought up the lack of scalability. For example, P8 stated: \textit{`When there are a lot of people around that have ordered something – and in the colour spectrum, there’s not heaps of colours that you could actually put on [an eHMI], it would be very hard to distinguish'}. 
For example, P8 suggested \textit{`another unique identifier'}, such as a \textit{`hologram'}, whereas P13 suggested a combination of \textit{`more expressive light patterns'}, such as \textit{`orange and purple [...] gently oscillat[ing] in the windscreen, so you could pick that that was your unique ride'}. This information should be also constantly available on the rider's personal devices, which P7 also considered as a limitation in the presented VR experience: \textit{`I think if I had the phone in my hand and I could reference the colour, that would have probably been helpful.'} \hlc{While the majority of participants (n=13) expressed concerns regarding the colour differentiation (i.e. blue and purple) for identifying their vehicle, only one participant (P13) expressed concerns about the abrupt yellow light to indicate alert, suggesting that \textit{`it’s just all a bit too much'} in reference to the number of different colours and animation patterns.}{adding information on perception of the yellow pattern (R1)}

\subsubsection{Pick-up Manoeuvre and Proximity to Rider}
More than half of the participants (n=8) commented on the vehicle's manoeuvre when picking up the other rider (Actor 2) or themselves and on the proximity of the vehicle towards the rider when coming to a stop. Opinions hereby varied widely; for example, two participants commented that they \textit{`were scared'} (P2) and \textit{`became really wary and alert'} (P10) when the vehicle was approaching them. P2 further commented that \textit{`she was just paying attention at the [vehicle's] sharp movement rather [than] the colour at that point'} and that \textit{`a taxi or a [manual] car would move towards a kerb with a smoother movement'}. While these two participants voiced the impression that the vehicle would almost run them over, 5 other participants expressed more positive perceptions. For example, P10 stated that the manoeuvre \textit{`was quite predictable'} and \textit{`you really feel like [the vehicle] is slowing down as it's approaching [and] there is no fear of the car coming at you'}. P5 even described a large gap between him and the vehicle once it had stopped: \textit{`[It] was really far away from me when it came to pick me up [...] the purple one. So, then I wasn't sure if it was coming to pick me up or if it was just stopping there for some reason. It seemed like I had to walk a few steps [...] It would have been more clear if it was closer to me at some sort of reasonable distance'}. However, he also added that he didn't know \textit{`what a reasonable distance would be'}, confirming \hl{the varying statements made by participants which suggest }that an optimal proximity depends on people's personal preference. In a similar vein, two other participants (P3, P13) emphasised the proximity of the stopped vehicle as the main cue to recognise their vehicle. P9 further added that this implicit cue raises expectations towards the SAV service: \textit{`If you have booked a destination in your [...] iPhone or whatever application it might be, and a vehicle turns up directly opposite you and you get in it, you would expect that vehicle to take you to that location.'}

\subsubsection{Additional Confirmation and Control}
While the vehicle's colour encoding and proximity towards the rider were considered important factors to gain confidence in identifying the correct car, participants also stated that they would need additional confirmation for a satisfactory customer experience with the SAV service. These comments were mostly related to the hypothetical boarding process, which was not covered in our scenario. For example, P7 said that while \textit{`the colour is really helpful from afar and getting prepared to get into a vehicle [...] there needs to be something a little bit more specific or unique to confirm'}. P3 who failed to recognise or correctly interpret the pulsing eHMI light pattern (L3) at the end of the last scene asked us: \textit{`How do you know it's safe to get in?'}. P5 suggested a \textit{``more verbal message, such as "Ready to board"'}. Relating to the safety driver in our scenario, P4 commented on the need for an additional confirmation from inside the car: \textit{`If there wasn’t another person in the shuttle, how can you ask if – or how can you confirm that it’s the right [vehicle]'}. Two participants related the need for additional cues also to the novelty factor of SAV services. P7 stated:  \textit{`There's going to be a while until I have full trust in something autonomous, so I need to have some kind of indication that I am getting into the right place and location'}. Similarly, P9 expressed that his trust towards the SAV service \textit{`would be built up on the number of times it does it correctly'}. While the need for additional unidirectional cues -- from the vehicle towards the rider -- were repeatedly mentioned, one participant (P7) also explicitly stated the need to gain some control over the vehicle. When asked about her repeated hand waving gesture while experiencing the VR prototype, P7 urged that the aspect of sensing and responding \textit{`is part of this change to autonomous vehicles'} and that \textit{`she would like to know, that she is influencing something'}.

\subsubsection{Trust and Shared Space}
Regarding our scenario of SAVs commuting in shared spaces, the interviews revealed that the majority of participants trusted the vehicle in the sense that they considered the chance of an accident as rather low. Participants' trust was induced by observing the vehicle's interactions with other pedestrians (n=9), including implicit cues (i.e. vehicle physically slowing down), and explicit cues (i.e. \hl{awareness} light pattern, L4), as well the low speed of the vehicle (n=3). Participants P3 and P10 further referred to the slow speed and small size of the vehicle in relation to the \textit{`very light pedestrian flow'}. Given this constraints, P3 even mentioned that \textit{`[he] would be very comfortable if it was driving a lot faster'} in the experienced context.

Interestingly, none of the participants objected the awareness light pattern (L4) -- often referred to as \textit{`alert'} signal -- in the shared space. Instead participants \textit{`[were] glad to see it turn a different colour'} (P7) in a potentially safety-critical situation and the vehicle being \textit{`really well lit up [...] to say that "I’m here and I see you'"} (P3). 
However, several participants urged that additional signage or segregation would be required to \textit{`let pedestrians know they are sharing the space with an autonomous vehicle, [...] because of safety reasons but also efficiency'}. P8 stated, similarly to P9, that \textit{`if you’re aware that something will always stop for you [...], you will just consciously not worry and will just do what you want to do'}. P2 further stated that \textit{`there wasn’t a marked difference between the roadway and where people were standing'}, which made her feel standing in the path of the vehicle. P6 stated similarly that \textit{`because there was no sign [...] for the vehicle to stop, it means it can stop anywhere'}, which made her \textit{`feel insecure'}. Instead she would expect the vehicle \textit{`to stop at a critical location'} in the shared space, such as a designated pick up area.

\section{Discussion}
In this study, we investigated the efficacy of eHMI communication in complex urban mobility scenarios exemplified through a ride-sharing service operating in an urban space shared by pedestrians and vehicles. Hereafter, we discuss the results according to the initial aims: the use of eHMIs to convey multiple messages simultaneously, pedestrians' perception of multiple AVs and their eHMIs, and AV-pedestrian interactions for SAVs in a shared space.

\subsection{Conveying multiple messages}
\hlc{The comprehensive literature review by \citet{Dey2020taming} found that there are no recommendations available at this stage regarding an eHMI's optimal information capacity (i.e. number of displays and number of messages), thus leaving it unclear for designers how to avoid potential cognitive overload.}{missing references and more focus on information capacity (R1)} Given the ride-sharing scenario, we designed the eHMI to display information that is relevant to an individual rider (i.e. identifying the vehicle) and the general public (i.e. status, intent and, awareness). \hl{Furthermore, we deliberately decided to display the information by a single display.} This meant that various messages were overlaid, namely the vehicle identifier encoded through colour with the vehicle's states and operations encoded through (animated) patterns. Further, given that various traffic situations were covered, distinct messages were displayed successively within a single display space. Participants, who experienced the scenario in VR, reported in the post-scenario interviews that they were mostly focusing on identifying their vehicle based on the colour encoding (n=10). Interestingly, however, the same number of participants also noticed and recalled the light pattern to signal awareness to other pedestrians, which they found important given the close proximity of the inattentive pedestrian in the represented situation. This may suggest that people filter for eHMI messages that are relevant to their particular goals or critical in terms of safety. \hlc{However, it also has to be noted that the sudden and clear change in colour (i.e. purple to yellow) and LED position (i.e. from only the bottom to all light bars lit up) was better distinguishable for participants. We therefore conclude that conveying multiple messages through a single low-res lighting display is possible to a certain extent, however, successful interpretation depends on various factors, including the respective situation and visual distinguishability of the different messages. Acknowledging the limitations of our study setup, we argue that more targeted investigations on eHMI's information capacity are needed, including such that compare different number of displays and messages, for different modalities and display types, and across different situations.}{discussion of information capacity (R1)}

\subsection{Perception of multiple eHMI-equipped SAVs}
\hl{Addressing the lack of use cases that investigate eHMI concepts beyond interactions with a single AV~\citep{Dey2020taming, Tran}}, we also wanted to test out if multiple SAVs in an urban area would impede comprehension of the eHMI. Specifically to our ride-sharing scenario, findings suggest that identifying a vehicle solely based on colour encoding has limitations when multiple SAVs commute through an area. Here, our findings point to the necessity of using a combination of colours or more unique light patterns; further, additional means for identifying a vehicle, e.g. through number plates, dynamic high-resolution displays or personal mobile devices, would improve riders' confidence in identifying their allocated SAV in high-traffic ride-sharing scenarios. However, our findings also show that users appreciate being able to identify their vehicle from a distance, which suggests that SAVs should adopt a combination of highly visible ambient eHMIs and additional cues for interactions in closer proximity. \hlc{Multimodal interaction concepts, such as Uber's light beacon in combination with their smartphone application~\citep{uberBeacon} or the additional use of haptic feedback for AV-pedestrian interaction~\citep{Mahadevan2018}, could be further adopted to the context of SAV ride-sharing.}{bridge to existing approaches (R1)}

In terms of prototyping and evaluating the efficacy of eHMIs in complex traffic scenarios, our approach \hl{of using 360-degree recordings} has limitations as we represented multiple vehicles by concatenating various recordings of a single vehicle. This was also emphasised by one participant who stated that \textit{`the lack of cars felt unnatural, [given it] was actually in a city like [anonymised for review]'} (P2). \hl{Furthermore, our prototyping setup did not allow participants to interact with the mobile SAV ride-sharing application during the VR experience. Thus, further work is needed to enhance the capabilities of 360-degree VR prototypes~\citep{Hoggenmueller2019hyper} for the design and evaluation of interactions with AVs and eHMIs.}

\subsection{AV-pedestrian interactions in shared spaces}
The majority of eHMI concepts has been designed for and evaluated in crossing situations on roads~\citep{Dey2020taming, Colley2020b}, whereas our study focused on interactions in shared spaces that are predominantly occupied by pedestrians. Generally, our participants, who experienced the scenario in VR, did not express any objections against sharing a pedestrianised area with autonomous vehicles; instead, some even stated that the SAV could have moved faster depending on the density of people. In terms of signalling awareness (L4), VR study participants appreciated a strong visual signal. This is interesting, as some of the participants from the expert workshops urged caution about strong alert signals when exploring the shared space scenario within the prototyping toolkit. Our findings also suggest aspects for further considerations, such as how to mitigate pedestrian behaviour that would cause an AV in a shared space to constantly come to a halt. Also, despite using eHMIs, additional information integrated into the immediate physical surroundings might still be needed, such as signs and road markings to indicate that an area is populated by AVs and to allocate dedicated stopping points for SAVs within a shared space.

\subsection{Additional implicit and explicit communication cues}
The results from our study also confirm the importance of implicit communication cues and, in that regard, extend previous findings~\citep{Rettenmaier2021, Risto2017,Moore2019} to the context of ride-sharing scenarios: indeed, more than half of our participants commented on the SAV's approaching manoeuvre and proximity to the rider in relation to trust and user experience. This is an interesting finding as it points out that implicit cues, such as motion and vehicle proximity, are not only relevant in safety-critical situations, such as crossing decisions, but also shape the user's experience with a service and need to be considered in the design. One VR study participant (P9) commented in this regard: \textit{`I think you can’t divorce the car displaying technology from that whole package. It has to be looked at holistically.'} Considering the eHMI only as one element within human-vehicle interaction design was also supported through some of our other findings. Participants commented that for our ride-sharing scenario additional communication channels, amongst others via personal devices and interfaces inside the vehicle, but also direct influence and control over the vehicle via sensor input is required. This highlights the need for future work to consider more carefully interaction trajectories and how interactions unfold involving a series of service touch points, as well as considering explicit and implicit human-machine interactions. Instead of only focusing on what information to communicate depending on the vehicle's proximity to a passenger, future frameworks should also consider the relationship of interaction modalities and the rider's spatial distance to the vehicle. We therefore propose to add another overarching dimension `implicit information' to the framework developed by~\citet{Owensby2018} in order to cover for the spatio-temporal vehicle movements. This would further emphasise that designing the vehicle's movement should not be left alone to engineers developing algorithms as it needs to be carefully designed to address trust and user experience towards SAV services more holistically.

\subsection{Limitations}
\hlc{Due to the exploratory nature of this study, the workshops and the VR evaluation study involved relatively small numbers of participants (14 people each). There was intentionally no overlap between the two groups of participants as having been part of the workshop would have influenced participant's knowledge and expectations in the VR study. However, this may have led to some of the contradictory observations; for example, in regards to preferences about the use of visual light signals. It thus remains unclear whether these observed differences stem from the background and characteristics (including participants' age) or the way participants assessed the scenes in the toolkit versus VR. Although we had a mix of participants in terms of their experience with VR, our sample was too small to identify whether and how this factor influences participants' perception of the SAV and its eHMI in VR. Furthermore, designing a comprehensive experience of a ride-sharing scenario, including multiple situations and eHMI messages, and following a design process, including several iterations and data collections, made it at times difficult to trace back findings to specific design decisions. These limitations point to questions that could be investigated in future studies and more targeted eHMI evaluations (e.g. re information capacity).}{Study limitations (R3)}

\section{Conclusion}
In this paper we presented insights from our human-centred design process and analysed participant interview data collected through a VR study involving a ride-sharing scenario recorded as a 360-degree VR prototype. 
While the light patterns we implemented were not necessarily identified as the ideal solution for the eHMI messages that an SAV should be equipped with, our study pointed out several suggestions for improvements, such as including cues in higher-resolution for close-proximity interaction and avoiding overriding existing norms (e.g. in regards to our pattern for pulling over). Importantly, beyond the specific light pattern design, we were able to uncover insights about the role of implicit (e.g. vehicle behaviour) and explicit (e.g. via the light pattern) cues. We found that participants filter for explicit cues that are either relevant to their goals or to ensuring the safety of pedestrians. Our study suggests that implicit cues, such as the way a vehicle approaches a waiting passenger, may be equally if not more important to `get right' in order to facilitate clear communication between SAVs and pedestrians. 

Our findings also offer insights on the design process and the value of using a staged prototyping approach. To that end, our toolkit catered for context-based eHMI design explorations in complex mobility scenarios at an early stage of the design process. However, design parameters beyond the eHMI (e.g. the AV's motion) were not captured in the toolkit representation. Recording staged scenarios through 360-degree video and evaluating these first-person interactions in VR yielded deeper insights about our eHMI design. We further found that immersive VR prototypes should support participants' use of personal devices in VR, such as smartphones, in order to allow for an evaluation of the holistic experience and the various service touch points of complex scenarios, such as ride-sharing. 

As physical driving behaviours seem to play a major role, not only in terms of pedestrian safety, but also passenger's experience with an SAV service, we further urge for more interdisciplinary collaborations between engineering and interaction design. 
We were in a unique position of having access to a real AV and working as part of a project team that included engineers as well as designers. Having to fully implement the light patterns and the autonomous behaviour of the SAV forced us to face technical constraints that may be overlooked in a wizard-of-Oz or computer-generated VR study~\citep{Tran}. For example, the limitations of the algorithms and cost map for creating more natural, human-like driving trajectories led to further investigations in regard to the vehicle's motion in pick-up scenarios. As a result, the robotic engineering team is implementing modifications to the actual path planning algorithm to imitate an S-curve pattern to cater for more intuitive human-machine interactions.

\bibliographystyle{frontiersinSCNS_ENG_HUMS} 
\bibliography{test}

\end{document}